# The Effect of Cosmic Ray Muon Momentum Measurement for Monitoring Shielded Special Nuclear Materials

J. Bae[1], S. Chatzidakis[1]

[1]School of Nuclear Engineering, Purdue University, West Lafayette

**ABSTRACT**

Recently, cosmic ray muons have been considered as a potential high energy radiation probe for monitoring and interrogation of dense, well-shielded special nuclear materials (SNM). Due to their high-penetrative nature, cosmic ray muons can easily penetrate shielded nuclear materials with minimal absorption and with leaving the target objects intact. However, despite the potential benefits from using cosmic ray muons for SNM monitoring, their widespread application has been limited for various reasons, including relatively low cosmic ray muon flux at sea level and the difficulty of measuring muon momentum in the field which can increase resolution and reduce measurement time. In this work, we explore in detail the effect of cosmic ray muon momentum measurement, focusing specifically on SNM monitoring applications. Three different types of SNMs (HEU, LEU, and Pu) surrounded by lead shielding with five different thicknesses (0, 5, 10, 20, and 30 cm), are analyzed using Monte Carlo simulation for three momentum measurement resolution levels (perfect, limited, and absent). $10^3$ muons were generated in the simulation which translates to 4 minutes of measurement time for a standard cargo container. We found that it is possible to identify and separate HEU, LEU, and Pu with high accuracy (> 3σ) when using muon momentum measurement (perfect and limited) even when 30 cm-thick lead shielding was used. Currently, it is not possible to identify or separate the SNMs with 30 cm thick lead shielding without muon momentum knowledge. Our results show that the effect of measuring muon momentum can be significant and can result in reduced measurement times by a factor of 3 to 4 and/or improved monitoring and imaging resolution.

## INTRODUCTION

A total of 3,689 confirmed malicious use and transportation incidents to smuggling high enriched uranium (HEU), plutonium (Pu), and other nuclear materials were reported since 1993 in a recent survey by the International Atomic Energy Agency (IAEA) [1]. Unlike conventional technologies currently used for prevention of unauthorized use of special nuclear materials (SNM), i.e., passive gamma, neutron, or NRF, cosmic ray muons have been considered as an unconventional radiation probe for monitoring of SNMs because muons are hardly absorbed by SNMs and the measured quantities of scattered angle and displacement can be estimated using multiple Coulomb scattering (MCS) approximation [2]. Most cosmic ray muon applications focus on radiography or monitoring using muons, e.g., spent nuclear fuel cask tomography [3], geotomography [4], and reactor monitoring or imaging [5]. However, extensive use of cosmic ray muons is mainly limited due to the relatively low cosmic ray muon intensity, approximately $10^4$ muons m$^{-2}$ min$^{-1}$ at sea level and the difficulty in measuring muon momentum in-situ. In this work, we explore the potential benefits of measuring muon momentum in addition to recording muon scattering angles for monitoring SNMs. Using Monte Carlo simulations, we perform three scenarios with different muon momentum knowledge levels: a) perfect muon momentum knowledge (e.g., from an ideal muon spectrometer), b) practical or limited muon momentum knowledge (e.g., limited resolution muon spectrometer), and c) no momentum knowledge (e.g., absence of muon spectrometer). In these scenarios, we simulate significant quantities of high enriched uranium (HEU), low enriched uranium (LEU), plutonium, and lead surrounded by lead shielding with thickness of 0, 5, 10, 20, and 30 cm. Scattering angle variance distributions of materials are constructed from stochastic muon transport simulations to analyze the performance of each scenario. $10^3$ to $10^4$ muon samples are simulated to gather enough statistics to scan a single standard cargo container with two-fold 3 m$^2$ muon detectors. Two levels of investigation are introduced: a) separation between SNM and non-SNM and b) identification of SNMs. For each scenario, three separation threshold levels are drawn between two neighboring variance distributions, i) Pb and Pu, ii) Pu and LEU, and iii) LEU and HEU, respectively. The receiver operating characteristic (ROC) curves are computed and the area under curve (AUC) is used as a metric to demonstrate the benefits of using a muon spectrometer.

## MONITORING SPECIAL NUCLEAR MATERIALS USING COSMIC RAY MUONS

A Cosmic ray muon is deflected due to multiple Coulomb scattering when traveling through SNMs and shielding materials. The expected scattering angle distribution depends on muon momentum and material properties. According to MCS approximation, the muon scattering angle distribution follows Gaussian distribution with zero mean as shown in Equation (1).

$$f(\theta|0, \sigma_\theta^2) = \frac{1}{\sqrt{2\pi}\sigma_\theta} \exp\left(-\frac{1}{2}\frac{\theta^2}{\sigma_\theta^2}\right) \qquad (1)$$

where $\theta$ is the muon scattering angle and $\sigma_\theta$ is a standard deviation of scattering angle. $\sigma_\theta$ is given by [6]:

$$\sigma_\theta = \frac{13.6\ MeV}{\beta c p} \sqrt{\frac{X}{X_0}} \left[1 + 0.088 \log\left(\frac{X}{X_0}\right)\right] \qquad (2)$$

where $\beta$ is the ratio of muon's speed to the speed of light, $p$ is the muon momentum, $c$ is the speed of light, $X$ is the thickness of the scattering medium, and $X_0$ is the radiation length. For example, when muons pass through uranium, a wider Gaussian distribution is expected than lead. Therefore, a hidden material could be identified by analyzing the muon scattering angle distribution if muon momentum is known or estimated. A typical scattering displacement measurement resolution is smaller than 1 mm [7], and a typical momentum measurement resolution is 0.5 GeV/c [8].

**DESCRIPTION OF CASE STUDIES**

Special Nuclear Materials with Lead Shielding

Significant quantities (SQ) of high-, low enriched uranium, and plutonium surrounded by lead shielding are considered in this case study. Main parameters of SNMs (HEU, LEU, and Pu) and Pb are summarized in Table 1. All SNMs and non-SNM are assumed to have a spherical shape surrounded by lead shielding thickness of 0, 5, 10, 20, and 30 cm (Figure 1). The effective radiation length of SNMs with a lead shielding layer is given by:

$$\frac{r_{obj}\rho_{obj}}{X_{0,obj}} = \frac{r_{SNM}\rho_{SNM}}{X_{0,SNM}} + \frac{L_{Pb}\rho_{Pb}}{X_{0,Pb}} \quad (SNM = HEU, LEU, and\ Pu) \quad (3)$$

$$r_{obj} = r_{SNM} + L_{Pb} \quad (4)$$

where $X_{0,i}$, $\rho_i$, and $r_i$ are radiation length, density, and radius of $i$ component where $i$ is SNMs, Pb, or both (object = SNM + Pb), and $L_{Pb}$ is the thickness of lead shielding. The radius of SNMs is determined based on their reported SQs. Total radius of object is the sum of SNM radius and lead shielding thickness. To compare the effective radiation length of SNMs with various thickness of lead shielding, radiation length number $R$ is defined as $X/X_0$. $X_0$ and $R$ as a function of thickness lead shielding, $L_{Pb}$ are shown in Figure 2. According to Equation (3), $X_{0,obj}$ is a function of a size and density. When the lead shielding thickness increases a total size increases whereas the density decreases. Counterbalancing effect by density and size is shown in thin lead thickness. To clarify the effect of lead shielding thickness, $X_0$ is replaced by $R$, which has a linear relation with a lead thickness.

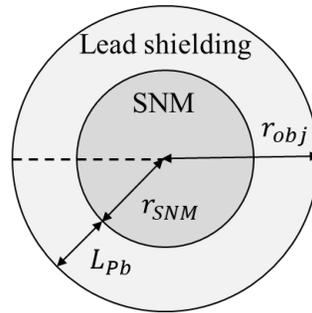

Figure 1. Cross section of spherical SNMs surrounded by a lead shielding.

Table 1. Density, radiation length ($X_0$), and significant quantity (SQ) of selected special nuclear materials and lead.

|  | LEU | HEU | Pu | Pb |
|---|---|---|---|---|
| Density [g/cm3] | 19.1 | 19.1 | 19.84 | 11.35 |
| $X_0$[cm] | 0.3166 | 0.3166 | 0.2989 | 0.5612 |
| SQ [9] | 75 kg | 25 kg | 8 kg | - |

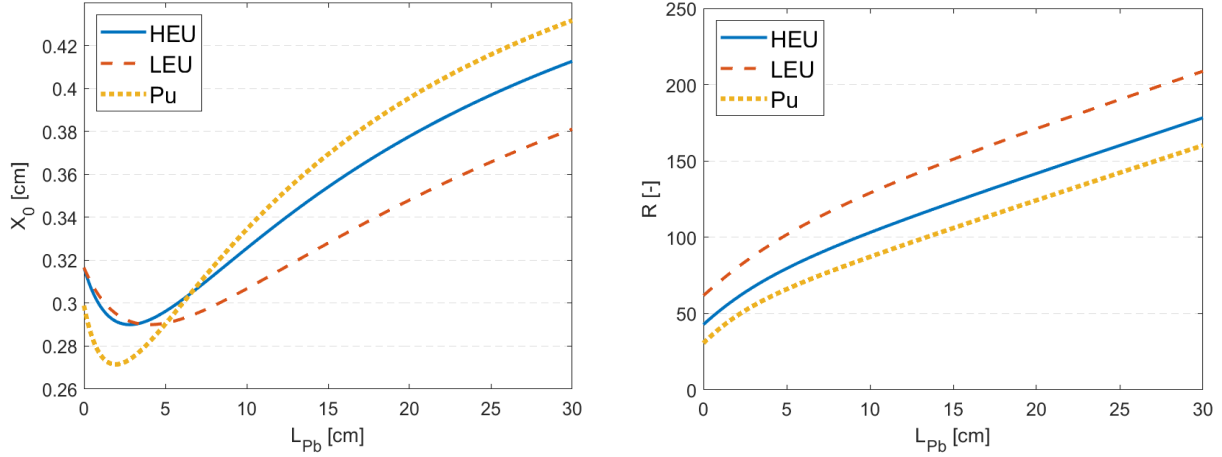

Figure 2. Effective radiation length (left) and radiation length number (right) of HEU, LEU, and Pu as a function of thickness of lead shielding.

Three Levels of Muon Momentum Knowledge

Three levels of muon momentum measurement resolutions are simulated a) perfect muon momentum knowledge from an ideal spectrometer, i.e., $\sigma_p = 0$ GeV/c, b) a limited momentum knowledge from a practical spectrometer with a momentum resolution of $\sigma_p = 0.5$ GeV/c, and c) unknown or not available muon momentum knowledge, i.e., p = 3.0 GeV/c for all muons irrespective of their true momentum. To generate cosmic ray muon momentum spectra, the MATLAB open source "Muon generator_v3" [10] is used and muon momentum is arbitrary selected within this spectrum to simulate unknown momentum knowledge.

Number of Muon Samples

Standard cargo container dimensions are 2.43 m wide, 2.59 m high, and 6.06 m in length. A vertical absolute integral intensity of muons at sea level is 72.5 $m^{-2}$ $s^{-1}$ $sr^{-1}$ [11]. Assuming two-fold 3 $m^2$ muon detectors with 3 m vertical distance, the expected muon count rate is approximately 22 $sec^{-1}$ per detector pair [12]. For a complete investigation of the entire volume of a standard container in 4 minutes, nearly 1,000 muons will be collected per a scanning volume. Increasing measurement time can improve SNM identification accuracy. However, if measurement time is limited, we need to measure muon momentum to maximize information availability for each incoming cosmic ray muon. The cases studied are summarized in Table 2.

Table 2. Summary of cases studied in the present work

| Case | $\sigma_p$ level | $N_\mu$ | SNMs | $L_{Pb}$ |
|---|---|---|---|---|
| I | Perfect | 1,000 | HEU, LEU, and Pu | 0, 5, 10, 20, 30 cm |
| II | Limited | 1,000 | HEU, LEU, and Pu | 0, 5, 10, 20, 30 cm |
| III | Absence | 1,000 | HEU, LEU, and Pu | 0, 5, 10, 20, 30 cm |
| IV | Absence | < 10,000 | HEU, LEU, and Pu | 0, 5, 10, 20, 30 cm |

## MONTE CARLO SIMULATION

Scattering Angle Variance Distribution

Although the number of muon counts depends on specifications of muon detector, ~$10^3$ muons can approximately translate to a few minutes measurement time. Therefore, in Monte Carlo simulation, $N$ number of muons are arbitrary generated within the momentum spectrum and their scattering angle distributions are recorded. Then this simulation process is iterated for $M$ times to construct scattering angle variance distribution. Each material has a characteristic variance distribution. The degree of separation between two variance distributions represents the system identification capability. In this simulation, characteristic Gaussian distributions are constructed for every muon after they pass through the SNMs and Pb based on muon momentum and material property. The muon scattering angle distribution $S(x_i)$ for $N$ muons is developed:

$$\sum_{i=1}^{N} f_i(x_i | 0, \sigma_{\theta,i}^2) = S(x_i) \quad (6)$$

where $x_i$ and $\sigma^2_{\theta,i}$ are the scattering angle and scattering angle variance of $i^{th}$ Gaussian distribution. $S(x_i)$ can be approximated when sample size is large enough:

$$S(x_i) \sim S(x_i | 0, s^2) \text{ when } N \gg 1 \quad (7)$$

This is analogous to describing an undefined function using Fourier series. Integration of each expected Gaussian distribution, $f_i$ develops a $S(x_i)$. When we consider a perfect muon spectrometer, $N$ number of $f_i$ will be used to describe $S(x_i)$. On the other hand, one or a few (it depends on measurement resolution) $f_i$ will be used for the absent and limited muon spectrometer, respectively. The result of the muon scattering angle distribution is $D(x_i)$. Assuming $D(x_i)$ follows the Gaussian approximation,

$$D(x_i) \sim D(x_i | 0, s_D^2) \quad (8)$$

Here, we introduce one parameter which can represent the difference between the results of the muon scattering angle distribution and estimation. $\Delta s^2$ is the difference between variance of resulting distribution, $s^2_D$ and estimated distribution, $s^2$.

$$\Delta s^2 = |s_D^2 - s^2| \quad (9)$$

$$E[\Delta s^2] = \frac{1}{M}\sum_{j=1}^{M} \Delta s_j^2 = \mu \quad (10)$$

$$Var[\Delta s^2] = \frac{1}{M}\sum_{j=1}^{M} (\Delta s_j^2 - \mu)^2 = \sigma^2 \quad (11)$$

The mean and variance of $\Delta s^2$ increase when two distributions do not agree each other. Distribution of $\Delta s^2$ is expressed as:

$$X(\Delta s_j^2) \sim X(\Delta s_j^2 | \mu, \sigma^2) \text{ when } M \gg 1 \quad (12)$$

## RESULTS

Effect of Lead Shielding

We have considered 5 thicknesses of lead shielding, 0, 5, 10, 20, and 30 cm that surrounds SNMs. Figure 3 shows the simulation results for variance distributions of Pb, LEU, HEU, and Pu with 5 and 30 cm lead shielding thicknesses using practical momentum knowledge ($\sigma_p = 0.5$ GeV/c, N = $10^3$ and M = $10^3$). Without Pb shielding, SNMs are successfully separated from lead. However, scattering angle variance distributions overlap substantially when a 30 cm-thick Pb shielding is present and SNMs are no longer easily separated from Pb or one another. Material separation and identification resolution can be enhanced by increasing measurement time or improving momentum measurement resolution.

Effect of Measurement Time

To demonstrate the effect of measurement time, two scenarios are compared: a) $10^3$ muons without momentum knowledge and b) four times longer measurement time in Figure 4. None of materials including SNMs and Pb is able to be identified using $10^3$ muons without muon momentum knowledge (left) whereas Pb is clearly separated from SNMs with longer measurement time. However, the different types of SNMs are not easily identified.

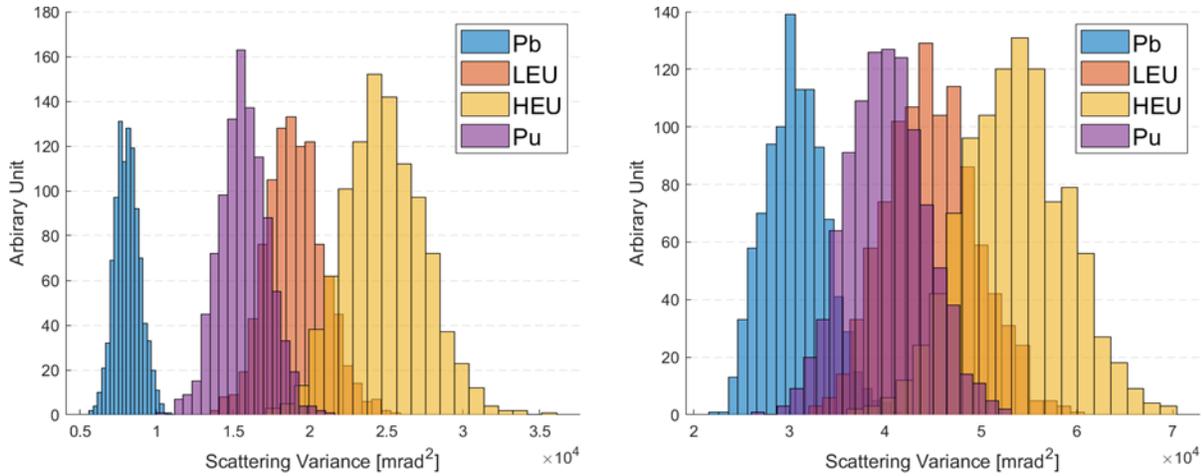

Figure 3. Scattering angle variance distributions of Pb, LEU, HEU, and Pu with a 5 cm (left) and 30 cm (right) lead shielding using N = 1,000 and limited muon momentum knowledge ($\sigma_p = 0.5$ GeV/c).

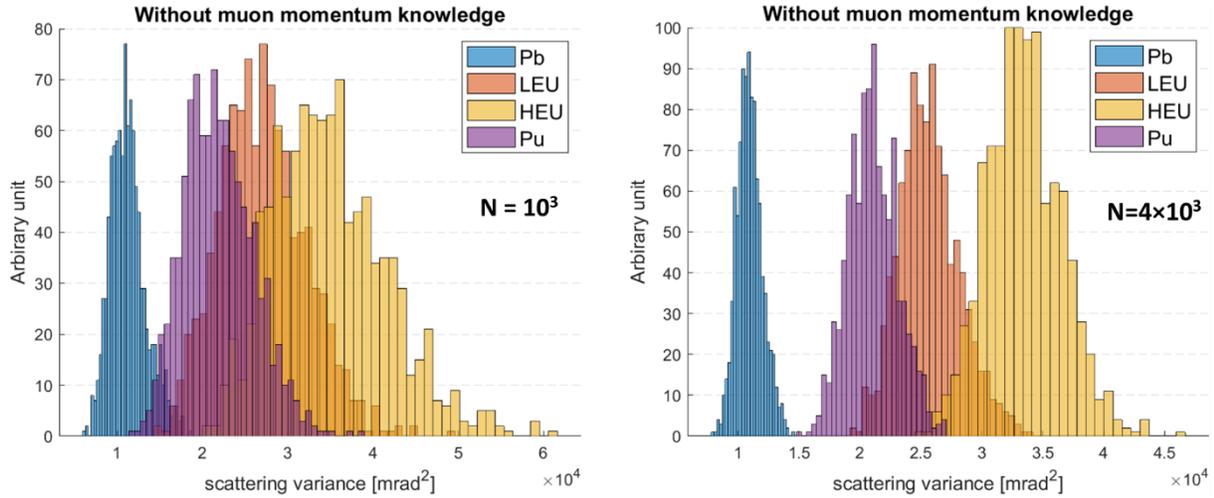

Figure 4. Variance distributions of Pb, LEU, HEU, and Pu with a 5 cm thick lead shielding for $10^3$ (left) and $4\times10^3$ muons (right) without muon momentum knowledge.

Separation and Identification of SNMs

To further quantify separation and identification of different materials, three threshold lines are drawn between two neighboring variance distributions. Pb-Pu, Pu-LEU, and LEU-HEU represent separation threshold lines for Pb and Pu, Pu and LEU, LEU and HEU, respectively. We only need three separation threshold lines because variance distributions of Pb, Pu, LEU, and HEU are placed in sequence under all conditions. Figure 5 shows the scattering angle variance distributions of SNMs with a 30 cm-thick lead shielding. Examples of three separation threshold lines are included as vertical dashed lines in Figure 5. ROC curves for Pb-Pu, Pu-LEU, and LEU-HEU are also shown in Figure 5. If muon momentum knowledge can be utilized in the SNM monitoring application, SNMs and non-SNM are easily separated and each SNM is clearly identified with high detection rate and low false alarm rate. On the other hand, without muon momentum knowledge, none of the materials can be distinguished without a high false alarm rate. Finally, in the more realistic situation with limited muon momentum knowledge, Pb separation from SNMs is improved although the false alarm will be somewhat increased.

Separation and identification AUCs as a function of lead shielding thickness are shown in Figure 6. For example, AUC of LEU-HEU drops from 1 to 0.78 without muon momentum knowledge whereas it only decreases to 0.89 with limited muon momentum knowledge. This highlights the advantage of measuring muon momentum, even when not perfect, in both SNM separation and identification in monitoring applications. To further quantify the benefit of measuring muon momentum, AUC curves as a function of number of muons for two SNM identifications with and limited muon momentum knowledge are plotted when they are surrounded by 30 cm lead shielding in Figure 7. In Figure 7, the intersection points of AUC curves and horizontal dashed line represent the number of muons required to achieve AUC = 0.9. For instance, to separate LEU and HEU with high confidence level (3σ) it would require 4,000 muons whereas only 1,000 are needed with limited muon momentum knowledge. For Pu and LEU separation, it requires 2,500 and 8,500 muons, respectively. Although the benefit of using muon momentum measurement in terms of necessary muon counts depends on various factors, we expect to shorten the measurement time by a factor of 3 to 4.

## CONCLUSIONS

Stochastic muon transportation simulations to generate muon scattering angle variance distributions by HEU, LEU, Pu, and Pb under three muon momentum knowledge levels with various lead shielding thickness conditions were performed. Without muon momentum level, it is impossible to separate SNMs (with 5 cm-thick lead shielding) from non-SNM material using $10^3$ muon samples (translated to a few minutes of monitoring time for a standard cargo container). With perfect muon momentum knowledge, SNMs and non-SNM are clearly separated and SNMs are identified. With limited muon momentum knowledge using a practical fieldable spectrometer, SNMs are distinguishable from non-SNM, however, each SNM is not certainly identified. To quantify the benefits of measuring muon momentum, despite a low resolution, ROC curves and AUC for three material threshold separation lines, Pb-Pu, Pu-LEU, and LEU-HEU were introduced. AUC quantity of LEU-HEU decreases from 1 to 0.78 without muon momentum knowledge whereas it stays stable by decreasing from 1 to 0.89 with a practical muon spectrometer. AUC curves for SNM separation lines as a function of number of muons were also used to compute required muon counts. To meet AUC = 0.9, about $10^3$ muons are required to specify HEU and LEU when they are shielded by 30 cm-thick lead with a practical muon spectrometer. On the other hand, $4\times10^3$ muons are necessary to meet AUC = 0.9 without muon momentum knowledge under identical conditions. Although the required number of muons for monitoring relies on various factor (i.e., AUC criteria, type of SNM separations, and shielding thickness), scanning time could potentially be shortened by a factor of 3 to 4 when measuring muon momentum with a measurement resolution of 0.5 GeV/c or better.

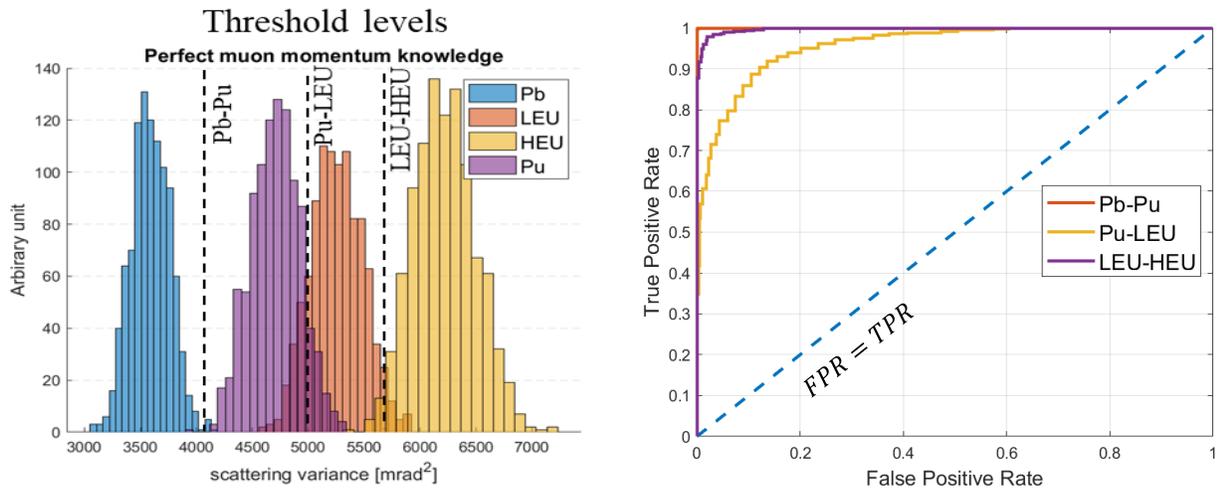

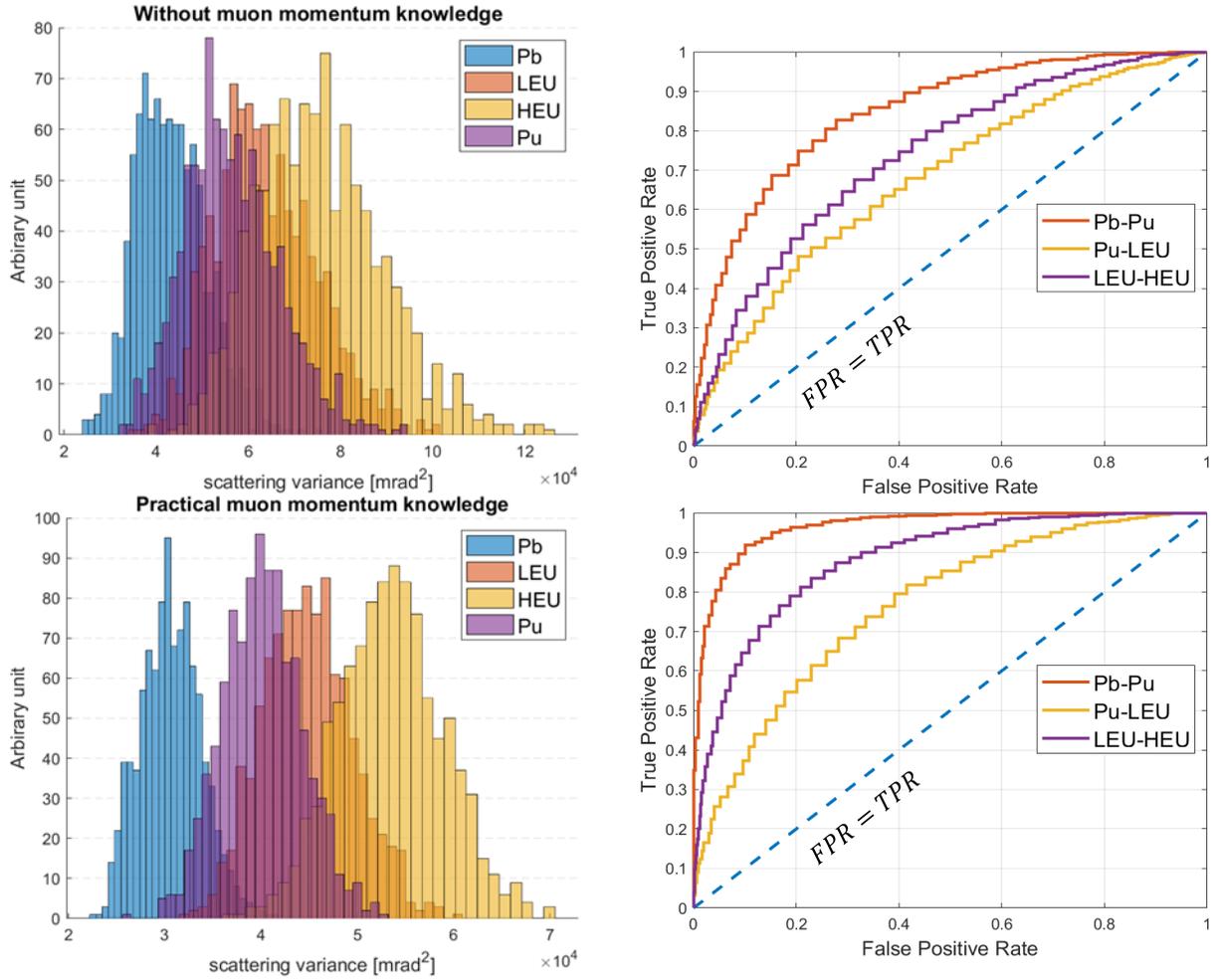

Figure 5. Muon scattering angle variance distributions of SNMs and Pb with a 30 cm thick lead shielding (left) and three ROC curves for each threshold level for two materials (right). Threshold levels for separation between two materials (Pb-Pu, Pu-LEU, and LEU-HEU) are also included.

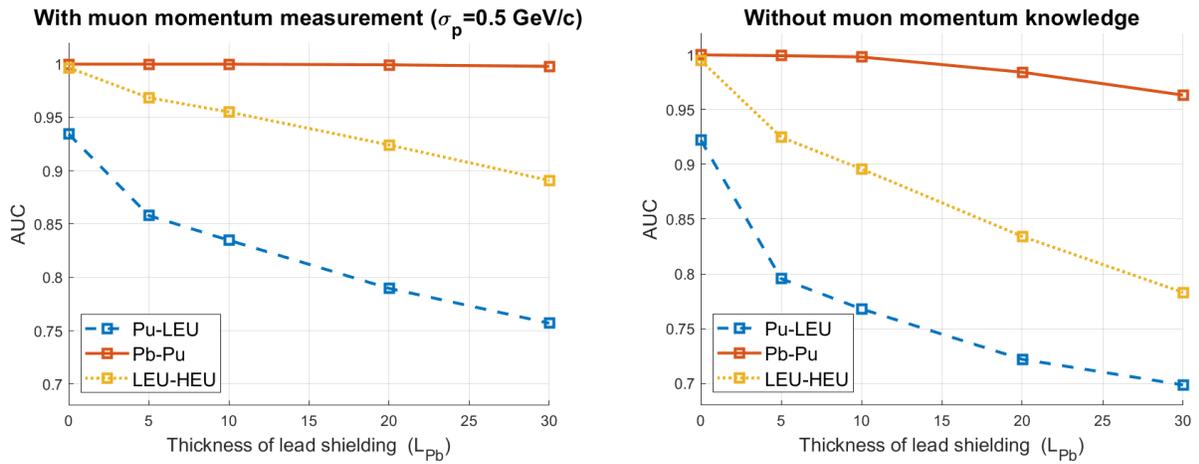

Figure 6. Pb-Pu, Pu-LEU, and LEU-HEU AUCs as a function of thickness of lead shielding without muon momentum knowledge (left) and with limited muon momentum knowledge (right).

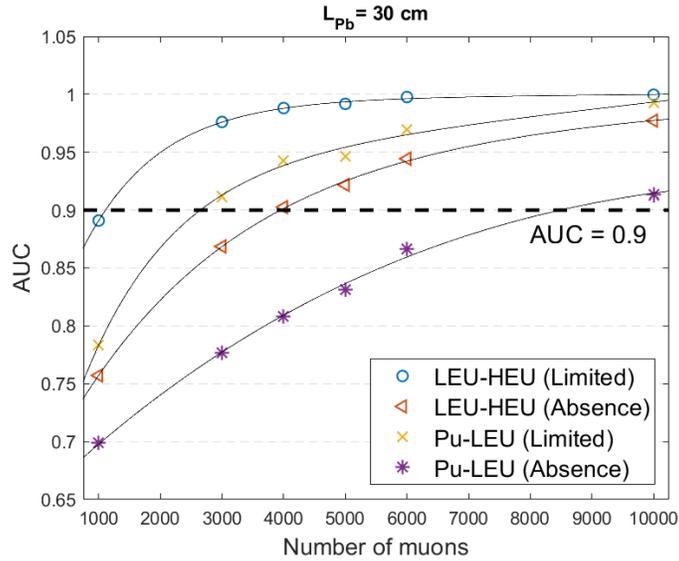

Figure 7. LEU-HEU and Pu-LEU separation AUC as a function of number of muons for limited and absent muon momentum knowledge with $L_{Pb}$ = 30 cm. The intersection points with curves and horizontal dashed line represent the number of muons required to achieve AUC = 0.9.

## ACKNOWLEDGMENTS

This research is being performed using funding from the Purdue College of Engineering and the School of Nuclear Engineering.

system for cargo containers," in *Journal of Instrumentation*, **9**, (2014).

[8]   J. Bae and S. Chatzidakis, "A Cosmic Ray Muon Spectrometer Using Pressurized Gaseous Cherenkov Radiators," *IEEE Nucl. Sci. Symp. Med. Imaging Conf.*, (2021).

[9]   IAEA, "IAEA Safeguard Glossary," (2001).

[10]  S. Chatzidakis, "Muon Generator.", [Online]. Available: https://www.mathworks.com/matlabcentral/fileexchange/51203-muon-generator (2021).

[11]  O. C. Allkofer, R. D. Andresen, and W. D. Dau, "The muon spectra near the geomagnetic equator," *Can. J. Phys.*, **46**, 301–305, (1968).

[12]  J. Bae, S. Chatzidakis, and R. Bean, "Effective Solid Angle Model and Monte Carlo Method: Improved Estimations to Measure Cosmic Muon Intensity at Sea Level in All Zenith Angles," *Int. Conf. Nucl. Eng.*, (2021).
11